\documentclass[12pt]{iopart}

\usepackage{graphicx}
\def\dbar{{\mathchar'26\mkern-12mu d}}
\usepackage{color}

\newcommand{\DyT}{Dy$_{2}$Ti$_{2}$O$_{7}$}

\newcommand{\HoT}{Ho$_{2}$Ti$_{2}$O$_{7}$}
\begin{document}

\title[]{Crystal Shape-Dependent Magnetic Susceptibility and Curie Law Crossover in the Spin Ices \DyT ~and \HoT} 

\author{L. Bovo$^1$, L. D. C. Jaubert$^2$, P. C. W. Holdsworth$^3$ and S. T. Bramwell$^1$ }

\address{1. London Centre for Nanotechnology and Department of Physics and Astronomy, University College London, 17-19 Gordon Street, London WC1H 0AJ, UK.}

\address{2. OIST Okinawa Institute of Science and Technology, Onna-son, Okinawa 904-0495, Japan.}

\address{3. Laboratoire de Physique, \'Ecole Normale Sup\'erieure de Lyon, Universit\'e de Lyon, CNRS, 46 All\'ee d'Italie, 69364 Lyon Cedex 07, France.}

\ead{l.bovo@ucl.ac.uk}

\begin{abstract}
We present an experimental determination of the isothermal magnetic susceptibility of the spin ice materials \DyT ~and \HoT ~in the temperature range $1.8 - 300$ ${\rm K}$. The use of spherical crystals has allowed the accurate correction for demagnetizing fields and allowed the true bulk isothermal susceptibility $\chi_{\rm T}(T)$ to be estimated. This has been compared to a theoretical expression based on a Husimi tree approximation to the spin ice model. Agreement between experiment and theory is excellent at {$T> 10$ ${\rm K}$}, but systematic deviations occur below that temperature. Our results largely resolve an apparent disagreement between neutron scattering and bulk measurements that has been previously noted. They also show that  the use of non-spherical crystals in magnetization studies of spin ice may introduce very significant systematic errors, although we note some interesting - and possibly new - systematics concerning the demagnetizing factor in cuboidal samples. Finally, our results show how experimental susceptibility measurements on spin ices may be used to extract the characteristic energy scale of the system and the corresponding chemical potential for  emergent magnetic monopoles. 

\end{abstract}

\maketitle

Spin ice materials~\cite{Harris,BramwellHarris,Ramirez,BramwellGingras} are generally discussed in terms of two approximate models: the `near neighbour' model introduced in Ref.~\cite{Harris,BramwellHarris} and the `dipolar model' introduced in Ref. \cite{Siddharthan,denHertog} and later refined in Ref.~\cite{Yavorskii}. The main difference between the two models is treatment of the magnetic dipole-dipole interaction that is truncated to near neighbour in the former model but retained in its full (classical) form in the latter~\cite{BramwellGingras}. The essential physics of spin ice is contained already in the near neighbour model, with the notable exception of magnetic Coulomb interaction~\cite{CMS} between emergent `monopole' defects~\cite{CMS,Ryzhkin,Jaubert1}. However it should be noted that quantum effects are neglected in both approaches and these may yet prove to be important at low temperature ($T <1$ ${\rm K}$)~\cite{Shannon,Benton}.  

The experimental magnetic susceptibility of spin ice materials is a property of considerable interest ~\cite{Harris,Bramwellchi,Snyder,Petrenko,Matsuhira,Quilliam}. It was first reported in Ref.~\cite{Harris,Bramwellchi} and analysed in terms of a Curie-Weiss law. Later theoretical developments~\cite{Ryzhkin,Isakov,Jaubert2} showed that the near neighbour model has a `Curie Law crossover' in the isothermal susceptibility:
\begin{equation}
\chi_{\rm T} = \frac{\mathcal{C}(T)}{T}.
\end{equation}
Here the effective Curie `constant' $\mathcal{C}$ crosses over from the paramagnetic value $\mathcal{C} = C$ 
($\approx 4$ ${\rm K}$ for \DyT~and \HoT) as $T\rightarrow \infty$, to exactly twice that number $\mathcal{C} = 2C$ in the low temperature limit. The detailed theoretical study of Ref.~\cite{Jaubert2} calculated the crossover analytically for a Husimi tree approximation to near neighbour spin ice. It showed that this crossover is an extraordinarily slow function of increasing temperature that is not complete until temperature of order $100$ $\times$ the characteristic energy scale $J$ of the interactions in the near neighbour model. This surprising result may be ascribed to the long range propagation of local constraints on spin configurations, leading to effectively amplified interactions that are captured in the infinite range summation of the Husumi tree approximation.

The predicted slow Curie Law crossover was found to agree very well with the experimental susceptibility of \HoT~measured at temperatures $T> 10$ ${\rm K}$ by bulk magnetometry and by neutron scattering at both at the Brillouin zone centre $002$ and the Brillouin zone boundary $001$~\cite{Jaubert2}. 
At lower temperatures the susceptibility measured by magnetometry still agreed with theory, as did that measured at the zone boundary, while that measured by neutron scattering at the zone centre showed systematic deviations. The present study was aimed in part at investigating this puzzling disagreement between bulk susceptibilities estimated by two different experimental methods. The resolution of this disagreement, presented here, will  facilitate the task of understanding the difference between experiment and theory, 
possible reasons for which are discussed in Ref.\cite{Jaubert2}.

The basic principle of magnetic thermodynamics is well known, but worth repeating here. We assume a sample that may be described in the cubic crystal system, as appropriate to spin ice. In this case the susceptibility tensor $\chi^{\alpha\beta}$ is diagonal with equal components (its representation quadric is a sphere). Hence, the isothermal susceptibility tensor may be replaced by a single scalar quantity, $\chi_{\rm T}$. If a magnetic field is applied to a sample of general shape by a solenoid then the incremental work done to change the magnetization of the sample is:
\begin{equation}
\dbar  W = \mu_0 \int \left[{\bf H}_0({\bf r}) + {\bf H}_1({\bf r}) \right]\cdot d{\bf M}({\bf r}) d{\bf r}, 
\end{equation}
where ${\bf r}$ is a position within the sample, ${\bf H}_0$ is the applied (solenoidal) field and ${\bf H}_1$ is the (irrotational) field arising from magnetic moment density within the sample. Homogeneity and parallel alignment of the magnetization and field are guaranteed only for ellipsoidal samples, in which case the above equation integrates to: 
\begin{equation}
\dbar  W = \mu_0 V{H}_{\rm int} d{M}, 
\end{equation}
where ${H}_{\rm int} = {H}_0 - \mathcal{D} {M}$ is the internal H-field, 
 $\mathcal{D}$ is the demagnetizing  factor, and ${M}$ and ${H}$ are the magnitudes of the vector quantities. For a spherical sample $\mathcal{D} = 1/3$. Assuming a linear medium in sufficiently small field the isothermal susceptibility $\chi_{\rm T}$ may be defined as : 
\begin{equation}
\chi_{\rm T} = \frac{M}{H_{\rm int}}.
\end{equation}

The earliest experimental measurements of the bulk susceptibility of single crystals spin ice~\cite{Harris,Bramwellchi} took careful account of the very strong demagnetizing fields~\cite{BramwellBook}. However the crystals used were not only flux grown and certainly more defective than more recently produced image furnace grown crystals, but also of an irregular block-like shape, rather than the ideal ellipsoid. These factors may have introduced systematic errors into the experimental estimation of the isothermal susceptibility of pristine spin ice. 

The work described in this paper used spherical crystals made by commercial hand-cutting of larger blocks which were provided by D. Prabhakaran~\cite{Prabhak}. Fig. 1 shows photographs of several such crystals, including at least one that is perfectly spherical to human perception and others less so. We measured the susceptibility of two nominally spherical crystals of each of \DyT ~and \HoT~in order to estimate the effect of departures from an ideal spherical shape. Magnetic susceptibility was measured using a Quantum Design SQUID magnetometer. Crystals were positioned in a cylindrical plastic tube to ensure a uniform magnetic environment. Three measurements were made on each sample: low field susceptibility (at $\mu_0H_0 = 0.005$, $0.01$ and $0.02$ ${\rm T}$) and field-cooled (FC) versus zero-field-cooled (ZFC) susceptibility. Also magnetic field sweeps at fixed temperature were performed in order to evaluate the susceptibility accurately and confirm the linear approximation.
The FC versus ZFC susceptibility measurements involved cooling the sample to $1.8$ ${\rm K}$ in zero field, applying the weak magnetic field, measuring the susceptibility whilst warming up to $300$ ${\rm K}$, cooling to $1.8$ ${\rm K}$ again and finally re-measuring the susceptibility while warming. Before switching the magnetic field off, field scans with small steps were performed in order to estimate the absolute susceptibilities. Furthermore, data taken at different fields were compared, such that the high temperature data (above $150$ ${\rm K}$) should correspond for each crystal. As previously reported~\cite{Bramwellchi}, no difference between field-cooled and zero-field cooled magnetization was observed down to $1.8$ ${\rm K}$ for both materials.

The measured susceptibility at high temperature was found to evidence the expected temperature-independent term $\chi_{\rm ti}$ that accounts for combined effects of diamagnetism and Van-Vleck susceptibility. The measured susceptibility displayed in this paper represents the Langevin contribution estimated by correcting the measured data for this term. The correction factors used were $\chi_{\rm ti}$(\DyT) $= 1.5 \times 10^{-3}$, $\chi_{\rm ti}$(\HoT) $= 1.2 \times 10^{-3}$. 

Fig. 2 shows our main result, where the experimentally determined susceptibilities in the form $\chi_{\rm T} T/C$ are compared with theory for coupling parameters appropriate to \DyT~and \HoT. Our theory is based on the Husimi tree method~\cite{Husimi} where calculations are done on a lattice made of corner-sharing tetrahedra, similar to a pyrochlore lattice in absence of any closed loops of spins beyond a tetrahedron~\cite{Jaubert2}. It is reminiscent of a Bethe lattice where each vertex would be a tetrahedron. The strength of this mean field method is to allow exact recursive results while correctly incorporating the correlations within each frustrated unit brick. It has been successfully compared to Monte Carlo simulations on the true 3-dimensional pyrochlore lattice~\cite{Jaubert3,Jaubert2}. From linear expansion in a field, one finds
\begin{eqnarray}
\chi_T=\frac{2 C}{T} \cdot\frac{1+e^{2J_{\rm eff}/T}}{2+e^{2 J_{\rm eff}/T}+e^{-6 J_{\rm eff}/T}},
\end{eqnarray}

%
%
where $J_{\rm eff}$ is the effective nearest neighbour coupling constant defined in~\cite{denHertog} and 
\begin{equation}\label{c}
C=\frac{\mathcal{N}\mu_{0}(g_{J}J \mu_{B})^{2}}{3 k}
\end{equation}  
Here $\mu_{0}$, $\mu_{B}$ and $k$ are, respectively, the vacuum permittivity, Bohr magneton and Boltzmann constant; $\mathcal{N}$ is the number of magnetic ions (R = Dy, Ho) per unit volume, $J$ is the the quantum number for total electron angular momentum of ${\rm R}^{3+}$ and $g_{J}$ is the Land\'e g-factor. 

A note about Eqn. \ref{c} is in order. The Langevin susceptibility for a {\it single} rare earth ion in a state defined by quantum number $J$ may be written:
\begin{equation}
\chi_{\rm T}^{\rm single} = \frac{\mathcal{N} \mu_0 g_J^2 \mu_B^2}{kT}\sum_{m_J} m_J^2 p(m_J)
\end{equation}
where $m_J$ is the magnetic quantum number and $p$ is the Boltzmann probability. In writing the Curie constant in the form of Eqn. \ref{c} it is assumed that the single ion crystal field ground state is a magnetic doublet composed entirely of the free ion states  $m_J = \pm J$  and that this doublet is well separated in energy from other states. In addition, an extra factor of 3 is necessary in the denominator of Eqn. \ref{c} to account for summing the single ion susceptibility tensor over four sets of ions in the pyrochlore structure of spin ice, each defined by a local trigonal quantisation axis belong to the cubic $\langle 111\rangle$ set. Another way to think about this factor of 3 is to consider that the single ion susceptibility has a single term in its susceptibility tensor while the cubic crystal has three equal diagonal terms. The factor 3 then guarantees that the trace of the tensor is invariant under rotations. 
Eqn. \ref{c} is reminiscent of the Curie constant of a free ion, $C=\mathcal{N}\mu_0g_J^2J (J+1) \mu_{B}^2/3 k$, on account of the restoration of symmetry as one passes from trigonal point symmetry to cubic space symmetry. The assumption that the ground state doublet is a pure $m_J = \pm J$ state is an approximation~\cite{Rosenkrantz, Cao, Malkin}, as symmetry requires admixtures of other $m_J$ states, but it is believed to be an accurate one~\cite{Rosenkrantz} and this is born out by our results. Note however, that while Fig. 2 has been obtained without any fitting parameter, in the comparison of Ref.\cite{Jaubert2} the magnetic moment of the rare-earth ion had to be adjusted by 4\% for comparison with bulk measurements.

Referring to Fig. 2, we may draw the following firm conclusions. First, different crystals of the same compound give essentially the same susceptibility $\chi_{\rm T}(T)$, showing that small departures from a perfect spherical shape are unimportant. Second, the experimental estimates of $\chi_{\rm T}$ for both spin ice compounds agree very closely with theory at $T> 10$ ${\rm K}$, but show systematic deviations at lower temperatures. Thus the Curie Law crossover reaches a maximum of $\mathcal{C} \approx 1.4 C$, significantly below the theoretical limit of $\mathcal{C}=2 C$. Third, the differences in the susceptibility between the two spin ice compounds accurately mirror the differences in their basic energy scales. In the standard dipolar spin ice model, the parameter $J_{\rm eff}$, which comprises dipole and exchange contributions, has been evaluated as $1.8$ ${\rm K}$ in Ho$_{2}$Ti$_{2}$O$_{7}$~\cite{Bramwell2001} and $1.1$ ${\rm K}$ in Dy$_{2}$Ti$_{2}$O$_{7}$~\cite{denHertog}. Ho$^{3+}$ and Dy$^{3+}$ having nearly the same magnetic moment and lattice constant, their dipolar interaction is fixed. Their nearest neighbour exchange coupling can vary though; if antiferromagnetic, it is responsible for making the creation of topological defects or ``monopoles'' less costly in energy. For the dumbbell model, in which the point dipoles of the dipole model are extended into needles which touch at the centres of the tetrehedral cells~\cite{CMS}, this translates into different values of the monopole chemical potential $\mu$. Creating a single monopole in the absence of intermonopole interactions costs $-\mu \approx 4.35, 5.7$ {\rm K} in \DyT~and \HoT~respectively, although the numerical result for the dipolar spin ice model is slightly higher that this ($4.6$ ${\rm K}$ for \DyT)~\cite{Jaubert1}. The parameters $J_{\rm eff}$ and $-\mu$ in the dumbbell model are related by~\cite{CMS}: 
\begin{equation}
-\mu = 2 J_{\rm eff} + \sqrt{2} C\left(4 \sqrt{6}-3\right)/6\pi.
\end{equation}
It is therefore clear from our results that susceptibility measurements on single crystal samples can be used as a general method of establishing either $J_{\rm eff}$ or $\mu$ in spin ice materials. 
 
The result, Fig. 2., for \HoT~differs from that reported in Ref.~\cite{Harris} and exhibited in Ref.~\cite{Jaubert2} in that it deviates from theory at temperatures below $\sim$ $10$ {\rm K}. It behaves similarly to the neutron scattering susceptibility at zero wavevector (see Ref.~\cite{Jaubert2}). Given plausible experimental uncertainties in the latter estimation we may conclude that there is is no longer a significant discrepancy between the zero wavevector neutron scattering susceptibility and the bulk susceptibility. 
However, this opens up a new puzzle: why did the  originally  measured bulk susceptibility agree so closely with theory down to a low temperature? 

To investigate this question we applied the measurement technique described above to four different cuboid crystals of \DyT~applying the standard demagnetizing correction for rectangular parallelepipeds~\cite{cube}. In particular, we chose a cubic[111] (nominally $\mathcal{D}_z = 1/3.0$), two squeezed [111] and [100] cuboidal shape (nominally $\mathcal{D}_z = 1/1.80$ and $\mathcal{D}_z = 1/2.0$, respectively) and an elongated [111] one (nominally $\mathcal{D}_z = 1/5.04$) where $[hkl]$ indicates the orientation of the crystal with respect of the applied field - representatives sketches of the crystals are shown in Fig. 3 panels a and b. The measured susceptibility in the form $\chi T/C$ is shown in Fig. 3, where a significant discrepancy with the result for the spheres is noted. The cuboid results are closer to the theory, but can also be projected onto the spherical result with effective demagnetizing factors of $\mathcal{D}_z = 1/3.20(1)$, $\mathcal{D}_z = 1/2.00(1)$, $\mathcal{D}_z = 1/2.17(1)$ and $\mathcal{D}_z = 1/6.10(1)$, respectively, or about $8-12$ $\%$ higher than the nominal ones. In view of this, it seems likely that the non-ellipsoidal shape of the sample used in Ref.~\cite{Harris} contributed to the agreement with theory reported in Ref.~\cite{Jaubert2}. 

The fact that the experimental data for the cuboidal crystals differs from that of the spherical crystals and also falls on the theory raises a number of issues. The demagnetizing correction used here is the `magnetometric' demagnetizing factor, which is spatially averaged over the whole cuboidal sample. If the sample fields were homogenous (as they are in a sphere), then the magnetometric demagnetizing factor would give the correct magnetostatic energy~\cite{Aharoni} and presumably a consistent equation of state and susceptibility. The observed insufficiency of the magnetometric demagnetizing factor implies that either the sample fields are not homogenous in the temperature range discussed, or that \DyT~exhibits shape dependent physics in this range, or a combination of these factors. In Table 1 we compare the magnetometric demagnetizing factor with the `ballistic' (or `fluxmetric') demagnetizing factor, which is found by averaging over a median plane in the sample~\cite{Joseph,Chen}, and with the `experimental' value needed to project the cuboidal data on to the spherical result, as described above. It is seen that the experimental value falls systematically in between the two theoretical estimates, suggesting that the method of averaging does play a role. It 
appears that all cuboidal samples studied here require a definite demagnetizing factor, which is neither magnetometric nor ballistic, suggesting a new systematic result for cuboids. 
It may seem surprising that a well-studied problem such as the demagnetizing factor could yield such a new result, but we note that spin ice offers a particularly tough test of the theory of demagnetizing factor as it combines extremely high magnetization density with negligible magnetic stiffness. On the other hand the fact that use of the magnetometric demagnetizing factor projects the susceptibility onto our theory is intriguing. One possible explanation is that the true susceptibility differs from the Husimi susceptibility because of local dipolar terms that be represented, in a mean field sense, as a correction to the demagnetising factor. However, we should also note that the effect of boundaries on the very low temperature susceptibility of spin ice has recently been discussed~\cite{Revell}. In the light of these observations it is not possible to decide at present whether the systematics described above apply to all magnets, or just to a restricted class that includes \DyT, or whether the projection of the cuboidal susceptibility onto theory contains any definite physics. It represents an interesting future challenge to decide between these possibilities. 

In conclusion, we have established the true bulk magnetic susceptibility of spin ice down to $1.8$ ${\rm K}$. There is a significant deviation from the Curie Law crossover expected on the basis of the Husimi tree approximation. In the future it will be useful to extend these results to yet lower temperatures and to compare them with more detailed microscopic models of spin ice that incorporate long range dipolar interactions~\cite{denHertog,Yavorskii} and quantum fluctuations~\cite{Shannon, Benton}. More generally we have shown that the use of spherical samples in magnetometry experiments on spin ice is generally highly desirable, to minimise systematic errors. This conclusion may well extend to several other rare earth magnets with the pyrochlore structure.

\ack
It is a pleasure to thank D. Prabhakaran and A. Boothroyd for supplying the crystals for this study and for related collaborations. This research was supported by EPSRC Grant EP/I034599/1. PCWH thanks the Institut Universitaire de France for financial support.

\newpage

\begin{figure}
 \begin{center}
 \renewcommand{\figurename}{Fig.}
\includegraphics[width=0.7\linewidth]{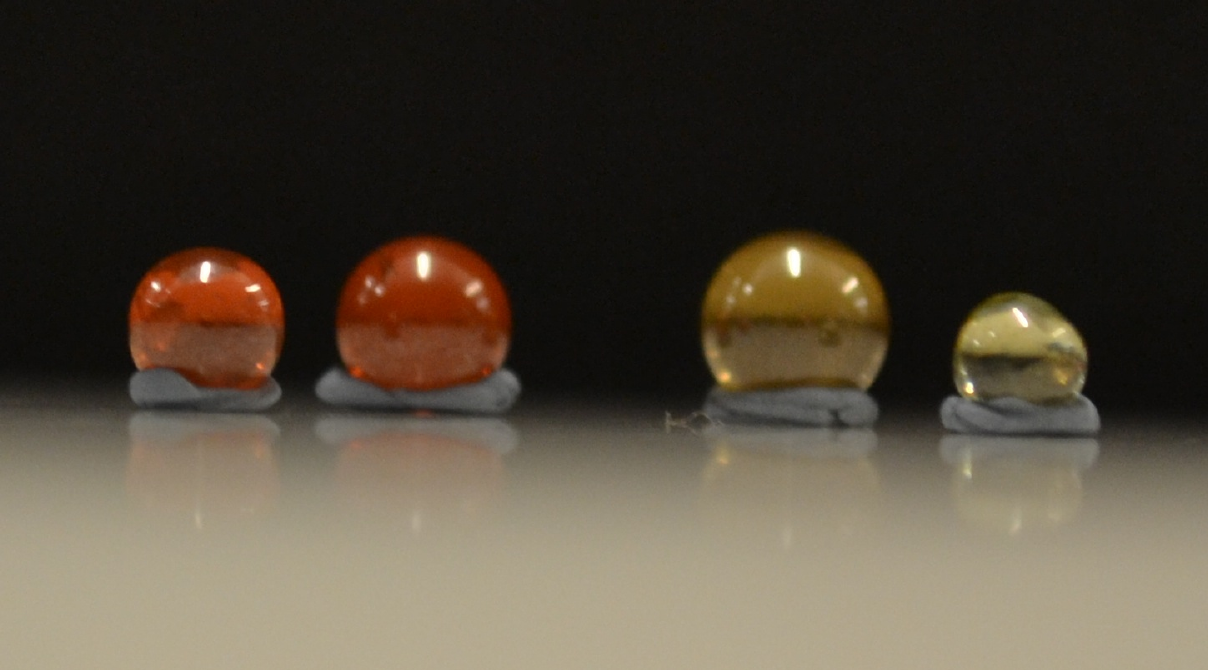}
\caption{Spin ice spherical crystals. From right to left \DyT~sample A and sample B; \HoT~sample C and sample D. Each sphere has a diameter ranging between $3$ to $4$ ${\rm mm}$.}
\end{center}
\end{figure}

\begin{figure}
 \begin{center}
 \renewcommand{\figurename}{Fig.}
\includegraphics[width=0.8\linewidth]{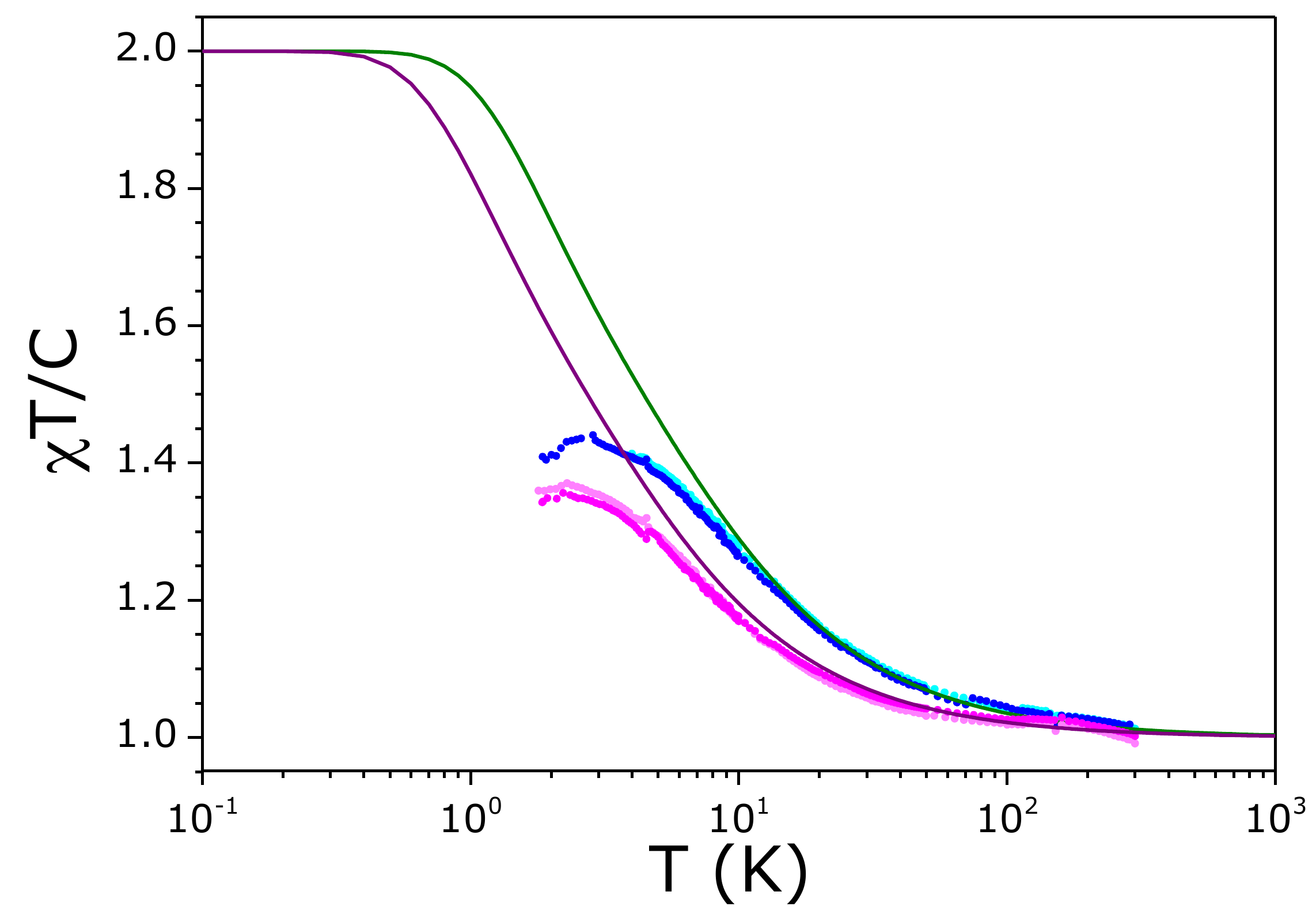}
\caption{Temperature dependence of the dimensionless quantity $\chi T/C$ calculated usung the expected $J_{\rm eff}$ (see~\cite{Jaubert2} and text for details) for \DyT ~(purple line) and \HoT ~(green line). The points represent bulk magnetometry data (see text for details) on spherical-shape single crystals of \DyT~ - sample A (magenta) and sample B (pink) - and of \HoT- sample C (blue) and sample D (light blue).}
\end{center}
\end{figure}

\begin{table} [h!]
\vspace{1cm}
\begin{center}
  \renewcommand{\arraystretch}{1.6}
    \begin{tabular}{ | c || c | c | c | c |}
    \hline
    Sample & $q=z/x$ & $\mathcal{D}_z$ ballistic & $\mathcal{D}_z$ magnetometric & $\mathcal{D}_z$ experimental \\ \hline
    \bf cubic[111] & 1 &0.2587 & 0.3333 & 0.312(1) \\ \hline
    \bf squeezed[100] & 0.5 &0.406 & 0.5050 & 0.461(2) \\ \hline
    \bf squeezed[111] & 0.4 & 0.5073 & 0.5482 & 0.500(3) \\ \hline
    \bf elongated[111] & 2 & 0.1109 & 0.1983 & 0.163(1) \\ \hline
    \end{tabular}
    \end{center}
    \caption{Numerical values of demagnetizing factors. $q$ denotes the aspect ratio of the sample length in the field direction to the orthogonal length ($z/x$). `Magnetometric' and `ballistic' estimates of $\mathcal{D}_z$ are taken from~\cite{Joseph,Chen}; the experimental $\mathcal{D}_z$ correspond to the corrected value used to project the experimental data onto the spherical result.}
    \label{table:ratio}
\end{table}
\vspace{1cm}

\begin{figure}
 \begin{center}
 \renewcommand{\figurename}{Fig.}
\includegraphics[width=\linewidth]{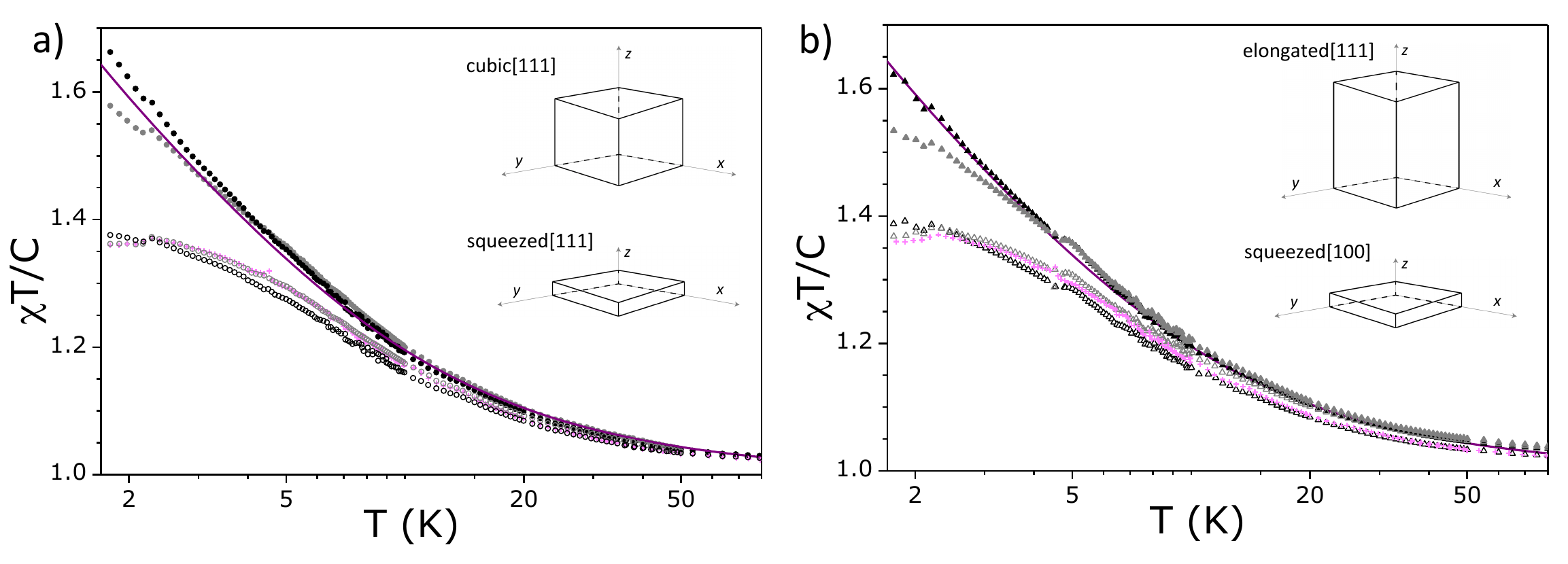}
\caption{Shape-dependent susceptibility in \DyT. Solid line (purple): calculated $\chi T/C$ (see~\cite{Jaubert2} and text for details). Bulk magnetometry data of {\bf a)} cubic[111] (black spots) and squeezed[111] (grey spots) crystals; {\bf b)} squeezed[100] (black triangles) and elongated[111] (grey triangles). Demagnetizing correction has been performed for cuboids according to Ref.~\cite{cube} (full symbols). Relatively small correction of about $10 \%$ (see text for details) can project the data (open symbols) onto the spherical result (pink crosses, corresponding to sample B).}
\end{center}
\end{figure}


\newpage
\begin{thebibliography}{99}


\bibitem{Harris} M. J. Harris, S. T. Bramwell, D. F. McMorrow, T. Zeiske and K. W. Godfrey, {\it Geometrical Frustration in the Ferromagnetic Pyrochlore {\rm Ho$_2$Ti$_2$O$_7$}}, Phys. Rev. Lett. {\bf 79}, 2554 (1997).

\bibitem{BramwellHarris} S. T. Bramwell and M. J. Harris, {\it Frustration in Ising-Type Spin Models on the Pyrochlore Lattice}, J. Phys.: Condens. Matter {\bf 10}, L215 (1998).

\bibitem{Ramirez} A. P. Ramirez, A. Hayashi, R. J. Cava, R. B. Siddharthan and B. S. Shastry, {\it Zero-Point Entropy in ÔSpin IceÕ}, Nature {\bf 399}, 333 (1999).

\bibitem{BramwellGingras} S. T. Bramwell and M. J. P. Gingras, {\it Spin Ice State in Frustrated Magnetic Pyrochlore Materials}, Science {\bf 294}, 1495 (2001).

\bibitem{Siddharthan} R. Siddharthan, B. S. Shastry, A. P. Ramirez, A. Hayashi, R. J. Cava and S. Rosenkranz, {\it Ising Pyrochlore Magnets: Low-Temperature Properties, "Ice Rules," and Beyond}, Phys. Rev. Lett. {\bf 83}, 1854 (1999). 
 
 \bibitem{denHertog} B. C. den Hertog and M. J. P. Gingras, {\it Dipolar Interactions and Origin of Spin Ice in Ising Pyrochlore Magnets},
 Phys. Rev. Lett. {\bf 84}, 3430 (2000). 

\bibitem{Yavorskii} T. Yavorskii, T. Fennell, M. J. P. Gingras, and S. T. Bramwell, \DyT ~{\it Spin Ice: A Test Case for Emergent Clusters in a Frustrated Magnet}, Phys. Rev. Lett. {\bf 101}, 037204 (2008). 

\bibitem{CMS} C. Castelnovo, R. Moessner and S. L. Sondhi, {\it Magnetic Monopoles in Spin Ice}, Nature {\bf 451}, 42 (2008).

\bibitem{Ryzhkin} I. A. Ryzhkin, {\it Magnetic Relaxation in Rare-Earth Pyrochlores}, J. Exp. and Theor. Phys {\bf 101}, 481 (2005).

\bibitem{Jaubert1} L. D. C. Jaubert and P. C. W. Holdsworth, {\it Signature of Magnetic Monopole and Dirac String Dynamics in Spin Ice}, Nature Phys. {\bf 5}, 258 (2009).

\bibitem{Shannon} N. Shannon, O. Sikora, F. Pollmann, K. Penc, and P. Fulde, {\it Quantum Ice: A Quantum Monte Carlo Study}, Phys. Rev. Lett. {\bf 108}, 067204 (2012).

\bibitem{Benton} O. Benton, O. Sikora and N. Shannon, {\it Seeing the Light : Experimental Signatures of Emergent Electromagnetism in a Quantum Spin Ice}, Phys. Rev. B {\bf 86}, 075154 (2012).

\bibitem{Bramwellchi} S. T. Bramwell, M. N. Field, M. J. Harris and P. I. Parkin, {\it Bulk Magnetization of the Heavy Rare Earth Titanate Pyrochlores - a Series of Model Frustrated Magnets}, J. Phys.: Condens. Matter {\bf 12}, 483 (2000). 

\bibitem{Snyder} J. Snyder, B. G. Ueland, J. S. Slusky, H. Karunadasa, R. J. Cava and P. Schiffer, {\it Low-Temperature Spin Freezing in the {\rm \DyT} Spin Ice}, Phys. Rev. B {\bf 69}, 064414 (2004).

\bibitem{Petrenko} O. A. Petrenko, M. R. Lees and G. Balakrishnan, {\it Titanium Pyrochlore Magnets: How Much Can be Learned From Magnetization Measurements?} J. Phys.: Condens. Matter {\bf 23}, 164218 (2011).

\bibitem{Matsuhira} K. Matsuhira, C. Paulsen, E. hotel, C. Sekine, Z. Hiroi and S. Takagi, {\it Spin Dynamics at Very Low Temperature in Spin Ice {\rm \DyT}}, J. Phys. Soc. Jap. {\bf 80}, 123711 (2011).

\bibitem{Quilliam} J. A. Quilliam, L. R. Yaraskavitch, H. A. Dabkowska, B. D. Gaulin, J. B. Kycia, {\it Dynamics of the Magnetic Susceptibility Deep in the Coulomb Phase of the Dipolar Spin Ice Material {\rm \HoT}}, Phys. Rev. B {\bf 83}, 094424 (2011).


\bibitem{Isakov} S. V. Isakov, K. S. Raman, R. Moessner and S. L. Sondhi,  {\it Magnetization Curve of Spin Ice in a {\rm [111]} Magnetic
Field}, Phys. Rev. B {\bf 70}, 104418 (2004).

\bibitem{Jaubert2} L. D. C. Jaubert, M. J. Harris, T. Fennell, R. G. Melko, S. T. Bramwell and P. C. W. Holdsworth, {\it Topological-Sector Fluctuations and Curie-Law Crossover in Spin Ice}, Phys. Rev. X {\bf 3}, 011014 (2013). 

\bibitem{BramwellBook} S. T. Bramwell, M. J. P. Gingras and P. C. W. Holdsworth, {\it Spin ice}
Book Editor(s): Diep, HT Source: {\it Frustrated Spin Systems}  Ed. Diep, H. T.,  pp. 367 - 456  World Scientific, Singapore, (2005). 

\bibitem{Prabhak} D. Prabhakaran and A. T. Boothroyd, {\it Crystal Growth of Spin-Ice Pyrochlores by the Floating-Zone Method}, J. Crys. Growth {\bf 318}, 1053 (2011). 

\bibitem{Husimi} K. Husimi. {\it Note on Mayer's theory of cluster integrals}, Journal of Chemical Physics {\bf 18}, 682 (1950).

\bibitem{Jaubert3} L. D. C. Jaubert, J. T. Chalker, P. C. W. Holdsworth and R. Moessner, {\it Three-Dimensional Kasteleyn Transition : Spin Ice in a {\rm [100]} Field}, Phys. Rev. Lett. {\bf 100}, 067207 (2008).

\bibitem{Bramwell2001}
S. T. Bramwell, M. J. Harris, B. C. den Hertog, M. J. P. Gingras, J. S. Gardner, D. F. McMorrow, A. R. Wildes, A. L. Cornelius, J. D. M. Champion, R. G. Melko and T. Fennell, {\it Spin Correlations in} \HoT~{\it : a Dipolar Spin Ice System}, Phys. Rev. Lett. {\bf 87}, 047205 (2001). 

\bibitem{Rosenkrantz}
S. Rosenkranz, A. P. Ramirez, A. Hayashi, R. J. Cava, R. Siddharthan and B. S. Shastry. {\it Crystal-Field Interaction in the Pyrochlore Magnet {\rm\HoT}}, J. Appl. Phys. {\bf  87}, 5914 (2000).

\bibitem{Cao}
H. Cao, A. Gukasov, I. Mirebeau, P. Bonville, C Decorse, and G. Dhalenne, {\it Ising Versus XY Anisotropy in Frustrated ${\rm R_2Ti_2O_7}$ Compounds as "Seen" by Polarized Neutrons}, Phys. Rev. Lett. {\bf  103}, 056402 (2009).

\bibitem{Malkin}
B. Z. Malkin, T. T. A. Lummen, P. H. M. van Loosdrecht, G. Dhalenne and A. R. Zakirov, {\it Static Magnetic Susceptibility, Crystal Field
and Exchange Interactions in Rare Earth Titanate Pyrochlores}, J. Phys.: Condens. Matter {\bf  22}, 276003 (2010).

\bibitem{cube} M. Sato and Y. Ishii, {\it Simple and Approximate Expressions of Demagnetizing Factors of Uniformly Magnetised Rectangular Rod and Cylinder}, J. Appl. Phys. {\bf 65}, 983 (1989).

\bibitem{Aharoni} Aharoni, A. 
{\it Introduction to the Theory of Ferromagnetism. }
Clarendon Press, Oxford, 1996. 

\bibitem{Joseph}Joseph, R. I.
{\it Demagnetizing factors in nonellipsoidal samples:  a review.}
Geophysics
{\bf 41} 1052 - 1054 (1976). 

\bibitem{Chen} Chen, D.-X., Pardo, E. and Sanchez, A.
{\it Demagnetizing factors for rectangular prisms.}
IEEE Trans. Magn. {\bf 41} 2077 - 2088 (2005). 

\bibitem{Revell}
 H. M. Revell, L. R. Yaraskavitch, J. D. Mason, K. A. Ross,	 H. M. L. Noad, H. A. Dabkowska, B. D. Gaulin,	 P. Henelius and J. B. Kycia, {\it Evidence of Impurity and Boundary Effects on Magnetic Monopole Dynamics in Spin Ice.} Nature Phys. {\bf 9}, 34 (2013).


\end{thebibliography}
\end{document}